# Machine Learning Assisted Model Reduction for Security-Constrained Unit Commitment

Arun Venkatesh Ramesh
*Student Member, IEEE*
Department of Electrical and Computer Engineering
University of Houston
Houston, TX, USA
aramesh4@uh.edu

Xingpeng Li
*Senior Member, IEEE*
Department of Electrical and Computer Engineering
University of Houston
Houston, TX, USA
xli82@uh.edu

***Abstract*— Security-constrained unit commitment (SCUC) is solved for power system day-ahead generation scheduling, which is a large-scale mixed-integer linear programming problem and is very computationally intensive. Model reduction of SCUC may bring significant time savings. In this work, a novel approach is proposed to effectively utilize machine learning (ML) to reduce the problem size of SCUC. An ML model using logistic regression (LR) algorithm is proposed and trained with historical nodal demand profiles and the respective commitment schedules. The ML outputs are processed and analyzed to reduce variables and constraints in SCUC. The proposed approach is validated on several standard test systems including IEEE 24-bus system, IEEE 73-bus system, IEEE 118-bus system, synthetic South Carolina 500-bus system and Polish 2383-bus system. Simulation results demonstrate that the use of the prediction from the proposed LR model in SCUC model reduction can substantially reduce the computing time while maintaining solution quality.**

## I. Introduction

The day-ahead generation scheduling in power systems is modeled as a complex optimization problem: security-constrained unit commitment (SCUC) [1]-[5]. It involves many physical and reliability constraints that ensure feasible and secure solutions while minimizing total system operational costs. However, as a mixed-integer linear programing (MILP) model that determines the commitment schedule of generators in day-ahead operations, SCUC is computationally intensive. Based on day-ahead solutions, in real-time, the solutions are utilized in solving a linear programming (LP) problem to perform economic dispatch in shorter time intervals. This is significantly faster compared to unit commitment due to the absence of binary constraints and reduced constraints. In practice, independent system operators (ISO) have limited time to solve the SCUC problem and post the solutions. Several literatures attempt to speed up the SCUC process by using techniques such as feasible region reduction, decomposition and/or heuristics [6]-[8]. A good warm-start solution also speeds up the process as seen in an example in [8]. However, identifying a good starting solution is hard; in addition, it does not reduce the model size.

Recently, machine learning (ML) methods have shown to be promising when used as a prediction or decision support mechanism in complex problems [9]-[13]. Combining ML techniques with traditional algorithms can improve the overall performance. Not only that, ML algorithms are robust to handle missing or noisy data, which can be a benefit for the sparse nature of complex bulk power system data during the learning process. Therefore, ML as a tool to learn the starting solution can be considered [14]-[15].

ML techniques and data-driven approaches have been utilized recently in aiding the SCUC process. Most papers predominantly focus only on enhancing network security [16]-[19]. In [16], a good starting solution was achieved for SCUC by integrating data-driven approach along with variable categorizing to improve the computational performance of SCUC. In [17], historical data was utilized to screen transmission constraints that are non-binding in the SCUC to speed up the process. Similarly, [18] uses an offline ML tool to learn about outage schedules and identifies planned outages. Similar to [17], [19] uses ML techniques to identify line outages under drastic weather conditions for SCUC to eliminate congested transmission constraints.

ML techniques were used to study the commitment schedules in [15] and [20]. A few machine learning techniques are proposed in [15] to use historical information to improve the performance of SCUC to solve identical instances in the future. However, [15] uses support vector method and k-nearest neighbor classification algorithm to learn warm-starting points for SCUC and yet are associated with drawbacks from infeasible problems. [20] utilizes an offline ML tool to categorize load profile into different categories with a pre-determined commitment schedule from history. However, [20] provides only a feasible solution and does not guarantee optimality or high solution quality.

Hence, we focus on building a supervised ML model to predict the commitment status of each generator $g$ in each time interval $t$ (24-hours) for day-ahead operations which are further aided by post-processes to determine the confidence of the solution. The commitment status of one implies the generator is ON whereas zero represents the generator is OFF. Ideally, a classification model can be utilized when the targets only belong to two classes, also known as binary classification. There are several classification models but only few models provide probabilities as an output. We also require a generative classifier that does not assume independence in pairs of input features. In this work, the input features are the respective normalized nodal demands and therefore cannot be considered independently in real-world data. Hence, the training model considered in this work is the logistic regression model which works ideally on Boolean outputs [21]-[22]. The proposed method in this paper focuses on innovative partial use of ML solutions in SCUC optimization to ensure high solution quality along with benefits of significant solve-time reductions.

This paper emphasizes that the commitment schedule data collected by ISOs can be utilized to train an ML model to improve grid operations and planning. The contributions of this work are presented as follows:

- A logistic regression-based ML model is proposed to use nodal demand profile as input to predict probabilistic generation commitment schedule.
- Two post-procedures are proposed to utilize the ML outputs to create a reduced-SCUC (R-SCUC) model while maintaining solution quality.

## II. Methodology

### A. SCUC Formulation

The objective of the SCUC is to minimize the operational cost of generators, $F(x,y)$ in (1), which includes the production, start-up and no-load costs. In (1), $x$ denotes the continuous variables of the problem such as generator dispatch points, line flows and bus phase angles; and $y$ denotes the commitment status and start-up binary variables. This is performed subject to generation, power flow constraints and reliability constraints in (2)-(3). The inequality constraints are modelled in (2) represent the minimum and maximum generation and transmission limits, the hourly generation ramp capability, and emergency 10-min reserve ramping capability while ensuring that reserves are held at the least to handle the failure of the largest generator. The equality constraints in (3) represent the nodal power balance and the power flow calculation. A detailed SCUC model can be obtained from our prior work [23].

*Objective:*

$$Min\ F(x,y) \tag{1}$$

*s.t.:*

$$G(x,y) \le b \tag{2}$$

$$H(x,y) = d \tag{3}$$

Since this work focuses on the ML based approach and its benefits, a simplistic SCUC model with only relevant base-case constraints are formulated. However, in future, a complete *N*-1 SCUC model can be utilized.

### B. Data Generation

To train ML models, a large amount of data is required. ISOs may possess historical data related to day-ahead operations, namely, commitment and demand schedules. These historical data would form a great starting point for practical real-world systems. For the test systems considered in this work, we have artificially created this dataset. By adding random noises in the standard nodal profile, several demand curves, $d_{i,g,t}$, were generated and their corresponding commitment schedule, $u_{i,g,t}$, can be obtained by solving the SCUC model described in sub-section II.A.

It can be noted that while trying to collect data for 1500 samples using a script, a few samples will result in infeasible load profiles which SCUC is infeasible for. This implies that the number of samples can vary for each system since infeasible load profiles are negated in the initially curated data. Therefore, approximately 1200-1500 samples denoted as $M$ are created for each test system, which can be considered as an equivalent of 3-4 years of data. The created $M$ samples, once shuffled, are split into two datasets: 80% training samples $M^{train}$ and 20% testing samples $M^{test}$.

### C. ML Model

The overall supervised ML approaches are described at a fundamental level in Fig. 1. The training and test samples are produced using data generation mentioned in sub-section II.B.

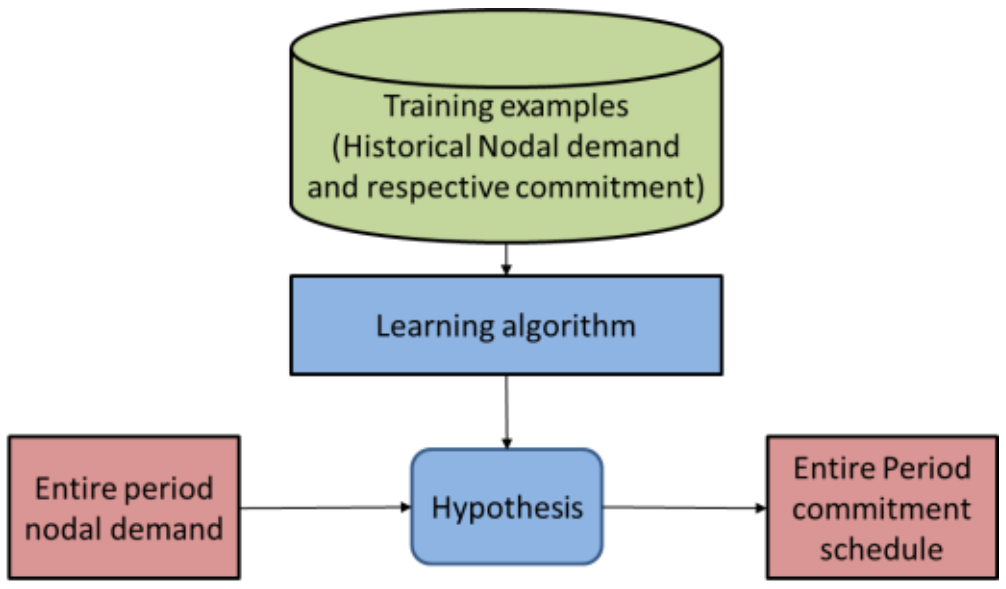

Fig. 1. Supervised ML approach.

The logistic regression (LR) model was chosen as the classifier method in this paper. Since LR is a well-established classifier method, several python packages provides a default package for outputs with input data. The LR model is implemented through scikit-learn package [24]. scikit-learn provides a package that fits the input and outputs based on the LR algorithm using the solver liblinear [25]. The liblinear solver uses a coordinate descent algorithm and only supports binary classification. The package is capable of handling one target or output at a time and the regularization is applied by default. This package trains the LR model with the following cost function in (4).

$$Cost = \frac{1}{2}w^T w - C\sum_{i=1}^{m} \log\left(\exp(-y_i(X_i^T w + c)) + 1\right) \tag{4}$$

where $w$ represents the trainable weights, $y_i$ represents the target/output, $X_i$ represents the input vector of sample $i$, and $C$ is the penalty.

In this work, the input features are the respective normalized nodal demands. The output targets are the generator $g$ commitment status which indicates the ON and OFF schedule in each time period $t$. A commitment status of 1 implies the generator $g$ is ON whereas 0 represents the generator $g$ is OFF. Therefore, the targets belong only to two different classes (binary classification).

The ML model accuracy can be verified using the post-processed outputs. Once the model is trained, the output probabilities, $P(u)_{i,g,t}^{ML}$, are post-processed to obtain the predicted commitment schedule $u_{i,g,t}^{ML}$: 1 if $P(u)_{i,g,t}^{ML} \ge P^{th}$, and 0 if $P(u)_{i,g,t}^{ML} < P^{th}$. $P^{th}$ is the probability threshold that varies between $0.2 \le P^{th} \le 0.8$. The accuracy in terms of $u_{i,g,t}^{ML}$, defined in (5), is calculated for both $i \in M^{train}$ and $i \in M^{test}$ using the optimal commitment $u_{i,g,t}$ obtained by solving SCUC as explained in sub-section II.B.

$$Acc = 1 - \frac{1}{m*N_g*N_t}\sum_{i=1}^{m}\left(\sum_{g\in G}\sum_{t\in T}\left|u_{i,g,t} - u_{i,g,t}^{ML}\right|\right) \quad (5)$$

where $m$ represents the number of samples, $N_g$ represents the number of generators, $N_t$ represents the number of time periods, $G$ represents the set of generators and $T$ represents the set of time periods.

### D. Proposed Methods

In this paper, the LR method described in sub-section II.C is extended with two proposed post-processing procedures. Once the LR model provides the probabilities of the generator commitment status, the decision boundary is then utilized to determine values of ML identified commitment status. The proposed methods in this section identifies which among the ML solution can further be processed to provide additional insights to reduce the complexity of the SCUC. The goal to choose the right post-procedure involves the elimination of infeasible problems while also maintaining the solution quality. The following are the proposed two procedures namely, *P1* and *P2*, to utilize the ML outputs to assist in establishing an R-SCUC for each power grid load profile of the testing samples:

- *P1*: R-SCUC where fix $u_{g,t} = 1$ if $u_{g,t}^{ML} = 1$ . The warm-start uses $u_{g,t} = 0$ if $u_{g,t}^{ML} = 0$.
- *P2*: R-SCUC where always ON/OFF generators are identified using $u_{g,t}^{ML}$ . For each testing sample (grid profile), if a generator $g$ is predicted to be always ON in 24-hour period then fix $u_{g,t} = 1$ for the entire 24-hour period for the corresponding generator. Similarly, if generator $g$ is always OFF in 24-hour period, then fix $u_{g,t} = 0$ for all periods for the corresponding generator. For all other generators, use warm-start $u_{g,t} = u_{g,t}^{ML}$.

For each sample $i \in M^{test}$, the above procedures *P1-P2* are implemented to perform R-SCUC and the quality of solution and time for computation are compared against *B1*. *B2* provides the maximum time reduction possible for each test sample. The overall flow of the proposed ML assisted SCUC process is represented in Algorithm 1.

**Algorithm 1** ML-assisted model reduction SCUC process

**For** $i \in M$
  randomize nodal demand
  Solve SCUC
  Store $d_{i,g,t}, u_{i,g,t}$, objective value and computing time
**End**
**Shuffle** $M$ samples
**Split** $M$ as 80% for $M^{train}$ and 20% for $M^{test}$
**Train** LR using $M^{train}$ for different $P^{th}$
Calculate training accuracy and test accuracy
**Tuning**: identify $P^{th}$ with best accuracy
Save ML predicted output probabilities
**For** $i \in M^{test}$
  Perform *B1-B2*, *P1-P2* and verify resultant SCUC for $d_{i,g,t}$
  record objective value and computing time
**End**

### E. Benchmark Methods

To obtain the boundary conditions, the benchmark methods are utilized to compare against the proposed methods. Therefore, two benchmark methods are utilized namely, *B1* and *B2*. The SCUC model represented in sub-section II.A is the benchmark model *B1* that does not utilize any ML information and is purely optimization. Whereas *B2* is an R-SCUC model only uses ML solution. This implies all the binary variables are fixed in SCUC, which is effectively converted into an economic dispatch problem. The following summarizes the benchmark methods:

- *B1*: normal SCUC that does not utilize any ML outputs $u_{g,t}^{ML}$, in which $u_{g,t}$ is solved only through MILP.
- *B2*: fix $u_{g,t} = u_{g,t}^{ML}$ and solve the reduced-SCUC (linear model in *B2*).

### F. Warm-Start vs Model reduction

A warm-start application provides starting point for the optimization solver to begin with, which may converge faster to the optimal solution. Note that the optimal solution could be very different from the starting point. Traditionally, most of the literature uses a warm-start application citing that ML models leads to infeasible solution [15], [19]-[20]. It is true that ML cannot purely replace optimization procedures since ML outputs are not 100% accurate. Therefore, utilizing ML outputs in completeness will result in infeasible problems. However, earlier research disregarded model reduction as a possible solution, i.e. partial usage of ML outputs.

In this paper, model reduction is proposed which fixes certain subset of variables/solutions that are determined from ML outputs with high confidence. This directly relates to reduction of variables and constraints in the MILP problem. Hence, the resultant MILP is an R-SCUC model which treats the remaining flexible generators as variables whereas the fixed generator status are treated like constants/parameters.

An advantage of warm-start is maintaining solution quality, i.e. the solution does not change with or without ML solution. However, the disadvantage is that time reduction is minimal in most cases. When compared against model reduction, R-SCUC results in significant time savings. Additionally through well-defined post-procedures to utilize the ML outputs, the solution quality is maintained to a high degree in R-SCUC.

## III. Results and Analysis

The SCUC mathematical model is implemented in AMPL. The data creation and verification steps are thus conducted using AMPL and solved using Gurobi solver with MIPGAP = 0.01. The ML step is implemented in Python 3.6. The computer with Intel® Xeon(R) W-2295 CPU @ 3.00GHz, 256 GB of RAM and NVIDIA Quadro RTX 8000, 48GB GPU was utilized. The proposed methods were validated with the following standard test systems summarized in Table I. Simulation results are presented in the following sub-sections.

TABLE I. Summary of Test Systems

| System | Gen cap (MW) | # gen | # branch |
|---|---|---|---|
| IEEE 24-bus [26] | 3,393 | 33 | 38 |
| IEEE 73-bus [26] | 10,215 | 99 | 117 |
| IEEE 118-bus [27] | 5,859 | 54 | 186 |
| South Carolina 500-bus [28] | 12,189 | 90 | 597 |
| Polish 2383-bus [27] | 30,053 | 327 | 2,895 |

### A. Decision Boundary Sensitivity Analysis

The decision boundary, $P^{th}$, is an important parameter that is utilized to classify generator status as ON or OFF. The outputs of LR algorithms are the probability of a generator *g* in time period *t* being ON. This does not affect warm start application since they are only used as starting values for the MILP.

However, in this paper, we focus on R-SCUC methods by directly using ML solution partially to reduce problem complexity. In the case of *B2*, benchmark method that uses complete ML solution, we notice $P^{th}$ significantly affects results. A key observation is that lower value of $P^{th}$ reduces the number of infeasible problems but affects solution quality (SQ) since more non-optimal generators are switched ON in certain time periods. The trade-off to consider in *B2* is more feasible problems vs solution quality. For example in the IEEE 24-Bus system, from Fig. 2, it can be seen that this changes from 48 (16.6%) infeasible test samples for $0.5 \leq P^{th} \leq 0.7$ to 85 (29.4%) infeasible test samples at $P^{th} = 0.8$. For $P^{th} = 0.9$ this increases to 261 (90.3%) infeasible test samples. The reason is that fewer generators are committed and they are unable to meet the load. Ideally, $0.5 \leq P^{th} \leq 0.7$ is the threshold based on the sensitivity analysis performed on *B2*.

This provides an important conclusion that ML solutions cannot completely replace optimization even if the accuracy is high. However, *B2* presents the maximum solve time (ST) savings achieved for feasible samples and this serves as a boundary condition to gauge the proposed methods, *P1* and *P2*.

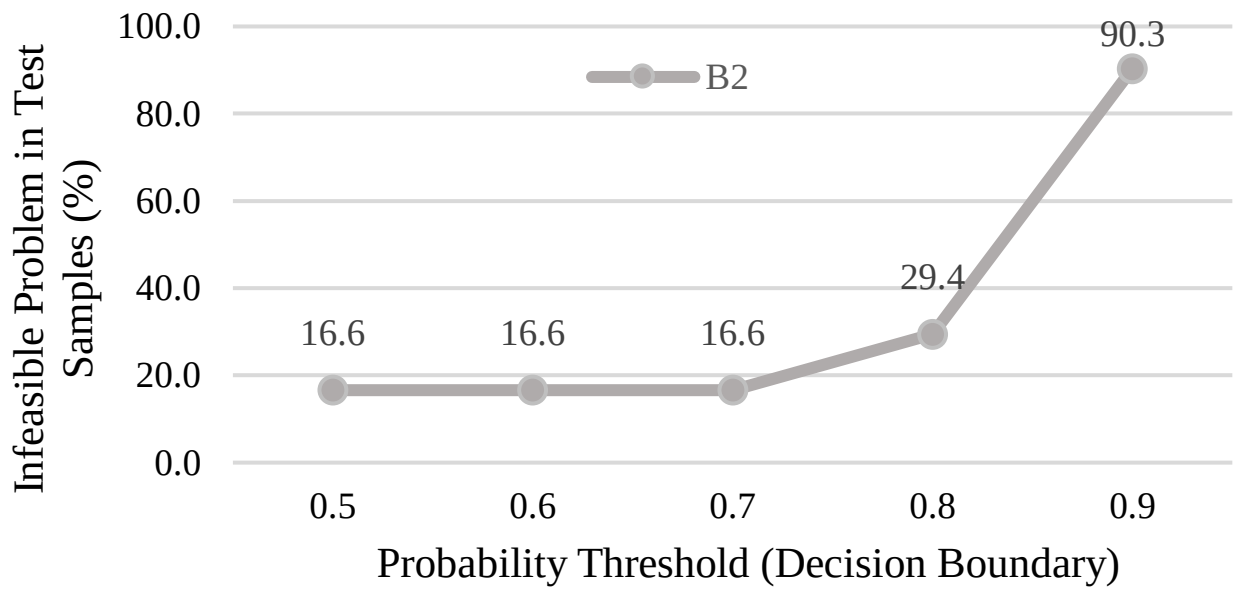

Fig. 2. Test sample infeasibility in IEEE 24-Bus system for *B2*.

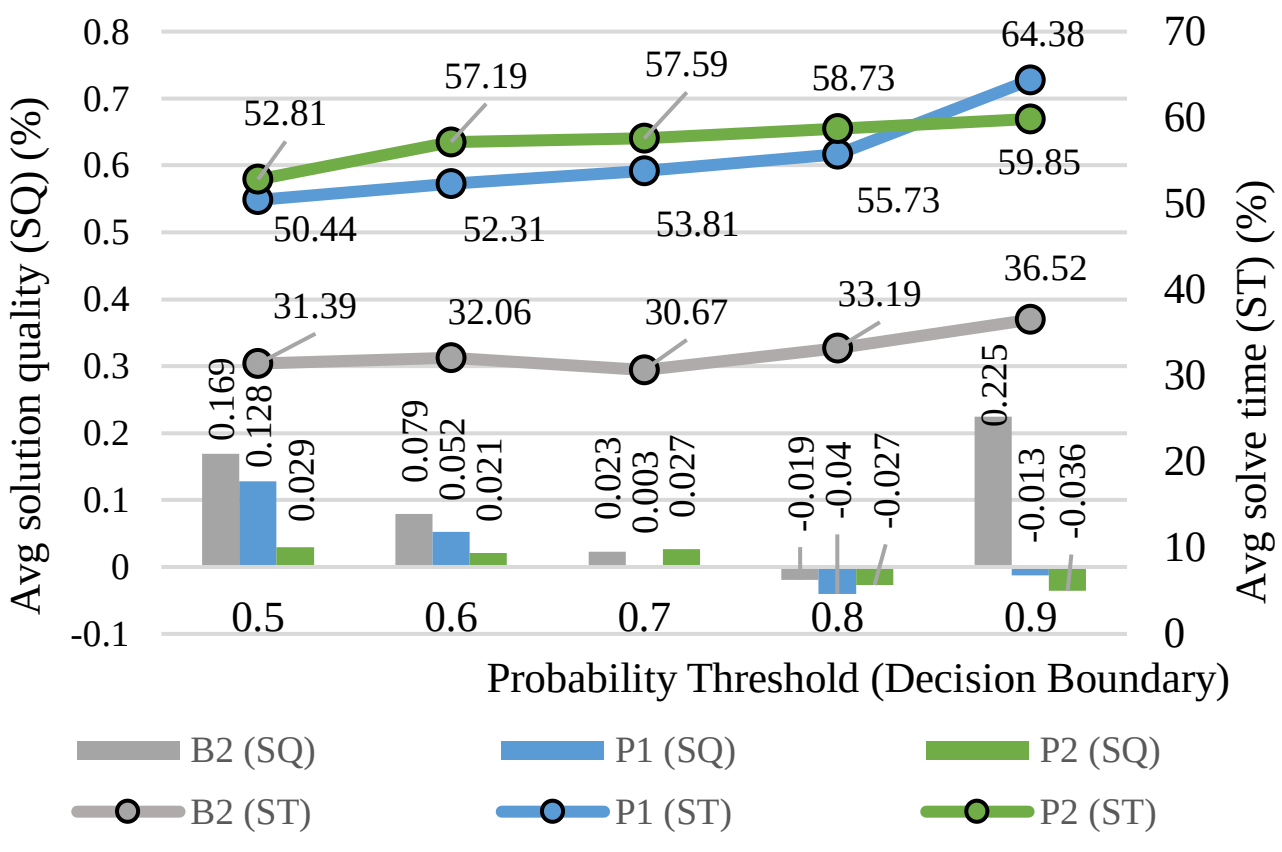

Fig. 3. B1-Normalized solution quality (SQ) and solve time (ST) averaged over test samples in percent for IEEE 24-Bus system with respect to decision boundary.

On further analysis, from Fig. 3, we can notice two trends for *P1* and *P2*. Firstly, as $P^{th}$ increases, the solution quality continuously improves. SQ values closer to 0 is more accurate as they represents the change in objective cost with respect to *B1*. SQ greater than 0 implies that the model provides a higher cost than *B1* and SQ less than 0 implies that the model provides lower cost than *B1*. This is because, (i) more variables are determined from optimization as opposed to ML solution and (ii) problem relaxations due to R-SCUC models are capable of achieving lower cost compared to SCUC model, *B1*. Secondly, time savings reduces as $P^{th}$ increases. Therefore, a trade-off between solution quality and time-savings is considered to obtain the best decision boundary.

### B. ML Training

Table II summarizes the $M$, $M^{train}$, $M^{test}$, accuracy, cumulative training time and decision boundary for each test system once the tuning and sensitivity analysis are completed. The LR algorithm provides high accuracy >93% for both the training and test samples for all test systems considered in this work.

The training time is the cumulative training time for all targets (each generator *g* in each time period *t*) for a particular test system. The training time is an offline step and is only implemented once for each system prior to the online step. Therefore, training time will not be a considered in R-SCUC which is online. Even so, training times are reasonable even for the large systems (Polish 2383-bus system) with ~85 min to train.

Ideally, the offline training can be done with grouped seasonal profiles that are similar in characteristics, patterns and resultant commitment schedules. This can increase the robustness of the algorithm. For the purpose of highlighting the benefits of LR algorithm and post-processing techniques, a higher variation in load profile is considered in this work. Here, it can be noted that a high accuracy directly implies a higher decision boundary, $P^{th}$, which implies ON and OFF generators are very well distinguished.

TABLE II. TRAINING SUMMARY

| # Bus | # Samples | | | Accuracy (%) | | Training time (min) | $P^{th}$ |
|---|---|---|---|---|---|---|---|
| | Total | Train | Test | Train | Test | | |
| 24 | 1,446 | 1,157 | 289 | 98.97 | 98.96 | <1 | 0.7 |
| 73 | 1,391 | 1,113 | 278 | 96.89 | 96.88 | <8 | 0.7 |
| 118 | 1500 | 1200 | 300 | 93.61 | 93.53 | <5 | 0.3 |
| 500 | 1499 | 1200 | 299 | 98.56 | 98.51 | <17 | 0.6 |
| 2383 | 1200 | 960 | 240 | 95.94 | 95.86 | <85 | 0.5 |

### C. Verification of the Proposed Method

Once each system was trained, a verification process was conducted for *P1* and *P2*. This was benchmarked against the *B1* method that does not use any ML solution and against the *B2* method that only uses ML solution to determine all generator commitment status. Therefore, the SQ from *B1* method is considered as 100% since it is purely an MILP optimization. The solutions cost and solution time of the R-SCUC models *P1*, *P2* and *B2* are represented as *B1*-normalized values.

Fig. 4 and Fig. 5 presents the *B1*-normalized objective value and computing time in percent when averaged over all test samples for each test system, respectively. From Fig. 4, *P2* provides comparable SQ to the *B1* method. However, *P1* leads to marginally increased costs since the ML solution may result in scheduling sub-optimal generators as ON. A key observation here is that not all samples of *B2* are feasible even though the accuracy is >93%. This is because *B2* fixes the status for all generators and only an optimal economic dispatch is then implemented. For the feasible samples, *B2* results only in marginal loss of SQ in IEEE 24-bus system (0.02%), IEEE 73-bus system (0.66%), IEEE 118-bus system (2.33%) and Polish system (2.02%) while it leads to a substantial increase of total cost on the IEEE 500-bus system. However, the SQ can be improved by *P1* and *P2* without infeasible problems. Therefore, the proposed methods, *P1* and *P2*, not only avoid infeasible problems but also maintain SQ. Among the proposed methods, *P2* offers the highest SQ similar to *B1* for all test systems.

From Fig. 5, though *B2* results in lower solution quality, it provides the most computational time savings ~95% in the Polish system since this eliminates all the binary variables in the problem. Therefore, the solution of *B2* serves as the maximum time-saving benchmark. *P1* and *P2* are R-SCUC models that reduce the number of variables and constraints and therefore result in considerable time-savings when compared to *B1*. Among the proposed methods, *P1* offers 50.9% time-savings and *P2* offers 38.8% time-savings on average across different test systems. *P1*, identifies more generator status (variables) to be ON, provides higher time-savings mainly because this procedure reduces more variables. In comparison, *P2* only identifies generators that are Always-ON or Always-OFF.

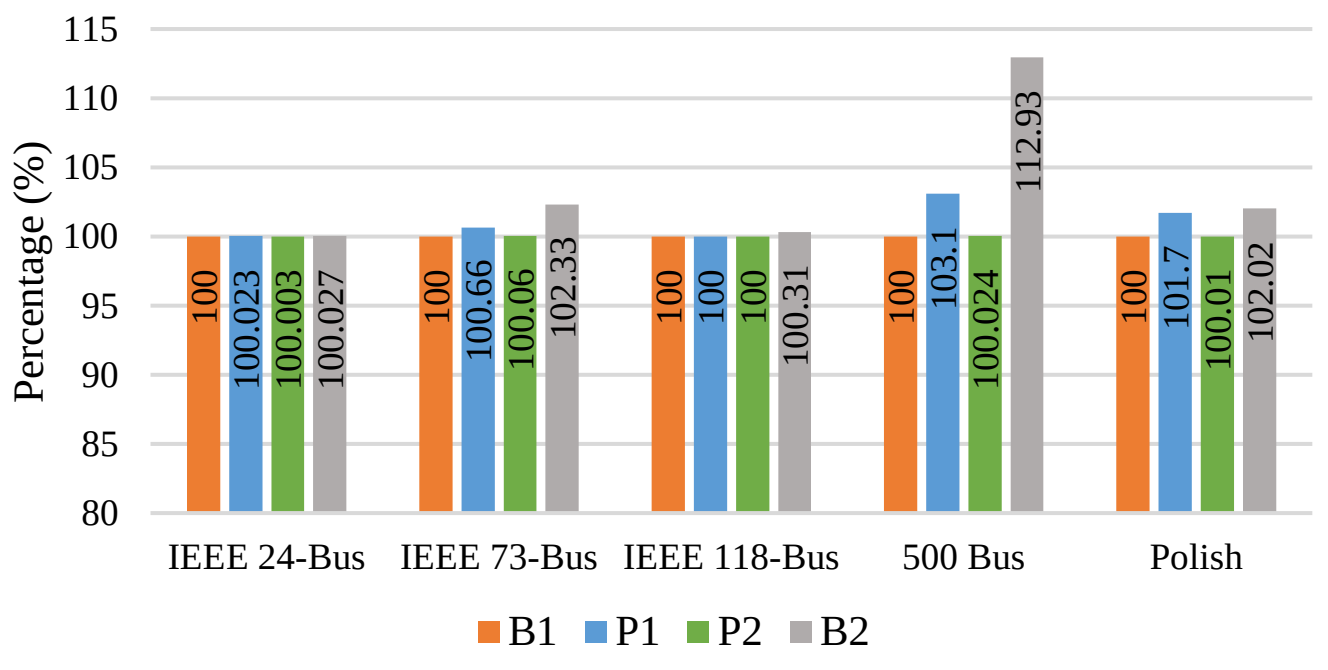

Fig. 4. Normalized objective value in percent averaged over test samples.

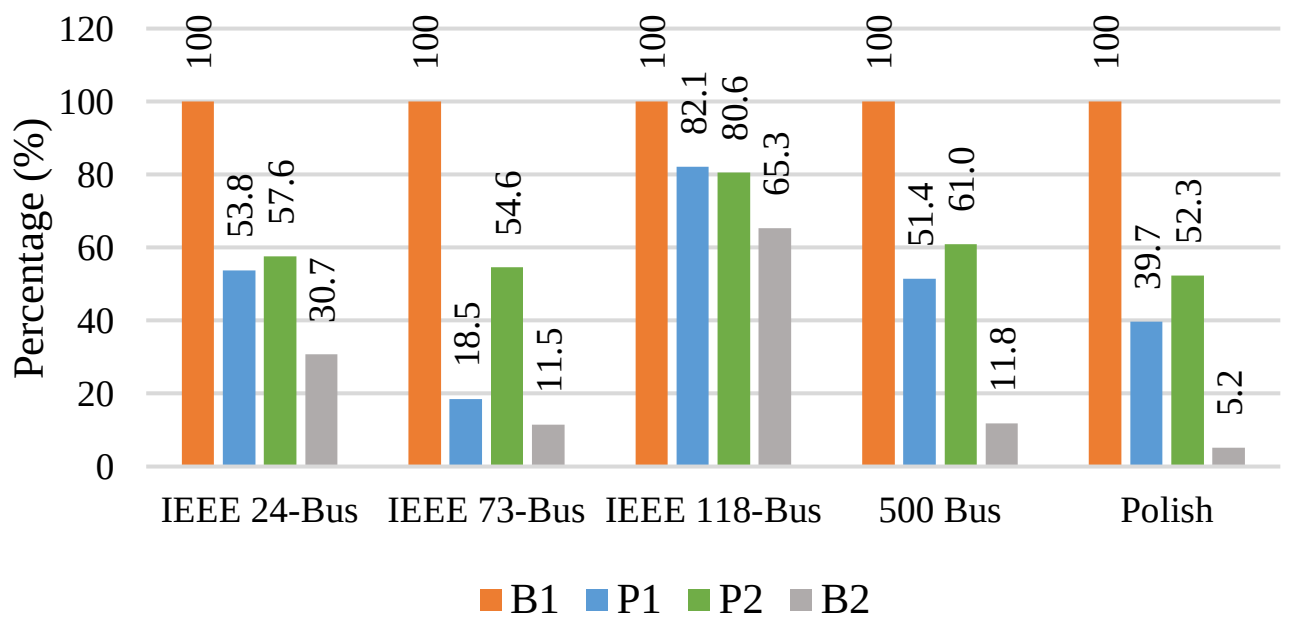

Fig. 5. Normalized computing-time in percent averaged over test samples.

### *D. Problem Size Reduction*

*B2*, *P1* and *P2* use ML solutions to fix the status of generators, which results in R-SCUC models. In particular, *B2* eliminates all binary variables in the problem whereas *P1* and *P2* results in decreased variables and constraints as seen in Table III. *B1* represents the pure optimization process and therefore has the full list of variables and constraints. A warm-start is not a proposed methodology in the paper. However, if a warm-start is utilized it will have the same problem size as the *B1* method since the ML solution is only used as a warm-start solution. By utilizing the ML solution, *B2*, *P1* and *P2* effectively reduces linear variables, binary variables, constraints and non-zeroes in the SCUC problem. As a result, it leads to smaller problem which results in time-savings.

TABLE III. AVERAGE PROBLEM SIZE FOR POLISH SYSTEM

| Procedure | Linear Var | Binary Var | Constraints | Non-zeroes |
|---|---|---|---|---|
| *B1* | 142,296 | 15,696 | 197,112 | 3,176,376 |
| *P1* | 128,856 | 3,700 | 153,296 | 2,924,834 |
| *P2* | 142,296 | 10,464 | 184,104 | 3,135,024 |
| *B2* | 125,330 | 0 | 136,168 | 2,027,850 |

## IV. CONCLUSIONS

It was evident that ML cannot directly replace the optimization procedure from *B2* since this lead to infeasible problems, which is detrimental. Not only that, the solution quality also suffers in feasible problems of *B2*. The only advantage is it points that almost 95% computational time can be reduced for large systems and is hence used as a benchmark boundary. The proposed ML assisted SCUC model reduction method can relieve the computational burden of the MILP problem while maintaining solution quality. The results also point that the ML model’s decision boundary selection plays a vital part in the model accuracy.

Further, once the models are trained, the proposed post-processing techniques, *P1* and *P2*, effectively utilize the ML predicted outputs without causing any SCUC infeasibility. The proposed approaches only use part of the ML solutions with high confidence to reduce the variables and constraints in SCUC. Not only that, the solution quality was not compromised especially in *P2*. *P1-P2* result in problem size reduction, which result in significant time-savings across multiple test systems. *P1* and *P2* result in time savings of 50.9% and 38.8%, respectively, on average across all the test systems while also resulting in high-quality solutions.

## V. FUTURE WORK

The future work would include the addition of renewable profiles and their impact on commitment schedule. A seasonal profile can be included to train different models for each distinct season. It can also be taken into account that in the same season, there may be only marginal differences day-to-day profiles or schedules, which will further improve the proposed ML-assisted SCUC method.

The proposed work in this paper considers all the SCUC constraints barring generator min-up and min-down time constraints. Modelling min-up and min-down time constraints may result in additional infeasibility of ML solutions. Hence, a post-processing technique to avoid infeasible solutions may be required. The proposed work mainly concentrated on generators that can be fixed ON with high confidence whereas future ML work can include the distinction of flexible and fixed generators. The proposed technique can also be compared against other ML models such as convoluted neural networks (CNN) and graph neural networks (GNN).

Predominantly, the focus has been on identifying commitment schedules with load profile patterns but future work can study other elements such as branch overloading to identify branch constraint relaxation or reduction using a similar technique.

## VI. References

[1] Mingjian Tuo and Xingpeng Li, “Security-Constrained Unit Commitment Considering Locational Frequency Stability in Low-Inertia Power Grids”, *arXiv:2110.11498*, Oct. 2021.

[2] Jin Lu and Xingpeng Li, “The Benefits of Hydrogen Energy Transmission and Conversion Systems to the Renewable Power Grids: Day-ahead Unit Commitment”, *arXiv:2206.14279*, Jun. 2022.

[3] Arun Venkatesh Ramesh, Xingpeng Li, "Network Reconfiguration Impact on Renewable Energy System and Energy Storage System in Day-Ahead Scheduling," *2021 IEEE Power & Energy Society General Meeting (PESGM)*, 2021, pp. 01-05, doi: 10.1109/PESGM46819.2021.9638033.

[4] Arun Venkatesh Ramesh, Xingpeng Li, "Enhancing System Flexibility through Corrective Demand Response in Security-Constrained Unit Commitment," *2020 52nd North American Power Symposium (NAPS)*, 2021, pp. 1-6, doi: 10.1109/NAPS50074.2021.9449717.

[5] Arun Venkatesh Ramesh, Xingpeng Li, "Reducing Congestion-Induced Renewable Curtailment with Corrective Network Reconfiguration in Day-Ahead Scheduling," *2020 IEEE Power & Energy Society General Meeting (PESGM)*, 2020, pp. 1-5, doi: 10.1109/PESGM41954.2020.9281399.

[6] Wu, L., & Shahidehpour, M. (2010). Accelerating the Benders decomposition for network-constrained unit commitment problems. *Energy Systems* (Berlin. Periodical), 1(3), 339–376.

[7] R. Saavedra, A. Street, and J. M. Arroyo, “Day-Ahead Contingency-Constrained Unit Commitment with Co-Optimized Post-Contingency Transmission Switching”, *IEEE Transactions on Power Systems*, 2020.

[8] Arun Venkatesh Ramesh, Xingpeng Li, and Kory W. Hedman, "An Accelerated-Decomposition Approach for Security-Constrained Unit Commitment With Corrective Network Reconfiguration," in IEEE Transactions on Power Systems, vol. 37, no. 2, pp. 887-900, March 2022, doi: 10.1109/TPWRS.2021.3098771.

[9] Cunzhi Zhao and Xingpeng Li, “Microgrid Day-Ahead Scheduling Considering Neural Network based Battery Degradation Model”, arXiv:2112.08418, Feb. 2022.

[10] Vasudharini Sridharan, Mingjian Tuo and Xingpeng Li, “Wholesale Electricity Price Forecasting using Integrated Long-term Recurrent Convolutional Network Model”, *arXiv:2112.13681*, Dec. 2021.

[11] Thuan Pham and Xingpeng Li, “Neural Network-based Power Flow Model”, *IEEE Green Technologies Conference*, Houston, TX, USA, Mar. 2022.

[12] Mingjian Tuo and Xingpeng Li, “Long-term Recurrent Convolutional Networks-based Inertia Estimation using Ambient Measurements”, *IEEE IAS Annual Meeting*, Detroit, MI, USA, Oct. 2022.

[13] Thuan Pham and Xingpeng Li, “Reduced Optimal Power Flow Using Graph Neural Network”, *arXiv:2206.13591*, Jun. 2022.

[14] V. Nair, S. Bartunov, F. Gimeno, I. von Glehn, P. Lichocki, I. Lobov, B. O'Donoghue, N. Sonnerat, C. Tjandraatmadja, P. Wang et al., "Solving mixed integer programs using neural networks", *arXiv preprint*, 2020.

[15] A.S. Xavier, F. Qiu, S. Ahmed, “Learning to solve large-scale security-constrained unit commitment problems,” *INFORMS Journal on Computing*, 2020.

[16] Y. Yang, X. Lu, and L Wu. “Integrated data-driven framework for fast SCUC calculation,” IET Generation, Transmission & Distribution, vol. 14, no. 24, pp. 5728-5738, Dec. 2020.

[17] S. Pineda, J. M. Morales, and A. Jiménez-Cordero, “Data-driven screening of network constraints for unit commitment,” IEEE Transactions on Power Systems, vol. 35, no. 5, pp. 3695-3705, Sept. 2020.

[18] G. Dalal, E. Gilboa, S. Mannor, and L. Wehenkel, “Unit commitment using nearest neighbor as a short-term proxy,” in Proc. 20th Power Syst. Comput. Conf., 2018, doi: 10.23919/PSCC.2018.8442516.

[19] F. Mohammadi, M. Sahraei-Ardakani, D. N. Trakas and N. D. Hatziargyriou, "Machine Learning Assisted Stochastic Unit Commitment During Hurricanes With Predictable Line Outages," in *IEEE Transactions on Power Systems*, vol. 36, no. 6, pp. 5131-5142, Nov. 2021.

[20] Y. Zhou, Q. Zhai, L. Wu and M. Shahidehpour, "A Data-Driven Variable Reduction Approach for Transmission-Constrained Unit Commitment of Large-Scale Systems," in *Journal of Modern Power Systems and Clean Energy*, doi: 10.35833/MPCE.2021.000382.

[21] Andrew NG’s machine learning course. Lecture on Supervised Learning [online]. Available: http://cs229.stanford.edu/materials.

[22] A.Y. Ng and M.I. Jordan, “On Discriminative vs. Generative Classifiers: A Comparison of Logistic Regression and Naive Bayes,” Advances in Neural Information Processing Systems, pp. 841-848, 2002.

[23] Arun Venkatesh Ramesh, Xingpeng Li, "Security Constrained Unit Commitment with Corrective Transmission Switching," *2019 North American Power Symposium (NAPS)*, Wichita, KS, USA, October 2019, pp. 1-6, doi: 10.1109/NAPS46351.2019.9000308.

[24] Pedregosa, Fabian, et al. "Scikit-learn: Machine learning in Python." *the Journal of machine Learning research* 12 (2011): 2825-2830.

[25] R.-E. Fan, K.-W. Chang, C.-J. Hsieh, X.-R. Wang, and C.-J. Lin. LIBLINEAR: A library for large linear classification Journal of Machine Learning Research 9(2008), 1871-1874.

[26] C. Grigg *et al*., "The IEEE Reliability Test System-1996. A report prepared by the Reliability Test System Task Force of the Application of Probability Methods Subcommittee," *IEEE Transactions on Power Systems*, vol. 14, no. 3, pp. 1010-1020, Aug. 1999.

[27] R. D. Zimmerman, C. E. Murillo-Sanchez, and R. J. Thomas, “MATPOWER: Steady-State Operations, Planning, and Analysis Tools for Power Systems Research and Education”, *IEEE Transactions on Power Systems*, vol. 26, no. 1, pp. 12-19, 2011.

[28] T. Xu; A. B. Birchfield; K. S. Shetye; T. J. Overbye,“Creation of synthetic electric grid models for transient stability studies,” accepted by 2017 IREP Symposium Bulk Power System Dynamics and Control, Espinho, Portugal, 2017.